\documentclass[aps,prb,preprint,amsfonts,showpacs,letterpaper]{revtex4-1}
\usepackage{graphicx, multirow,bm}
\usepackage{hyperref}

\begin{document}
%\begin{CJK*}{UTF8}{bsmi} 
	\title{Crystal field and magnetic structure of UO$_{2}$}
	\author{Fei Zhou}%(周非)}
	\author{Vidvuds Ozoli\c{n}\v{s}} 
	\affiliation{Department of Materials Science and Engineering, University of California, Los Angeles, CA 90095, USA} 
	\date{\today} 
	\pacs{71.27.+a, 71.15.Mb, 71.70.Ch} 
	\begin{abstract}
The properties of UO$_{2}$ result from rich $f$-electron physics, including electronic Coulomb interactions, spin-orbit and crystal field effects, as well as inter-ionic multipolar coupling. We present a comprehensive theoretical study of the electronic structure of UO$_{2}$ using a combined application of self-consistent DFT+$U$ calculations and a model Hamiltonian. The $\Gamma_{5}$ ground state of U$^{4+}$ and the energies of  crystal field excitations $\Gamma_{5} \rightarrow \Gamma_{3,4,1}$ are reproduced in very good agreement with experiment. We also investigate competing non-collinear magnetic structures and confirm 3-k as the $T=0$ K ground state magnetic structure of UO$_2$.
\end{abstract}
	\maketitle
%\end{CJK*}

\section{Introduction}
\begin{figure}[htbp]
\includegraphics[width=0.95 \linewidth]{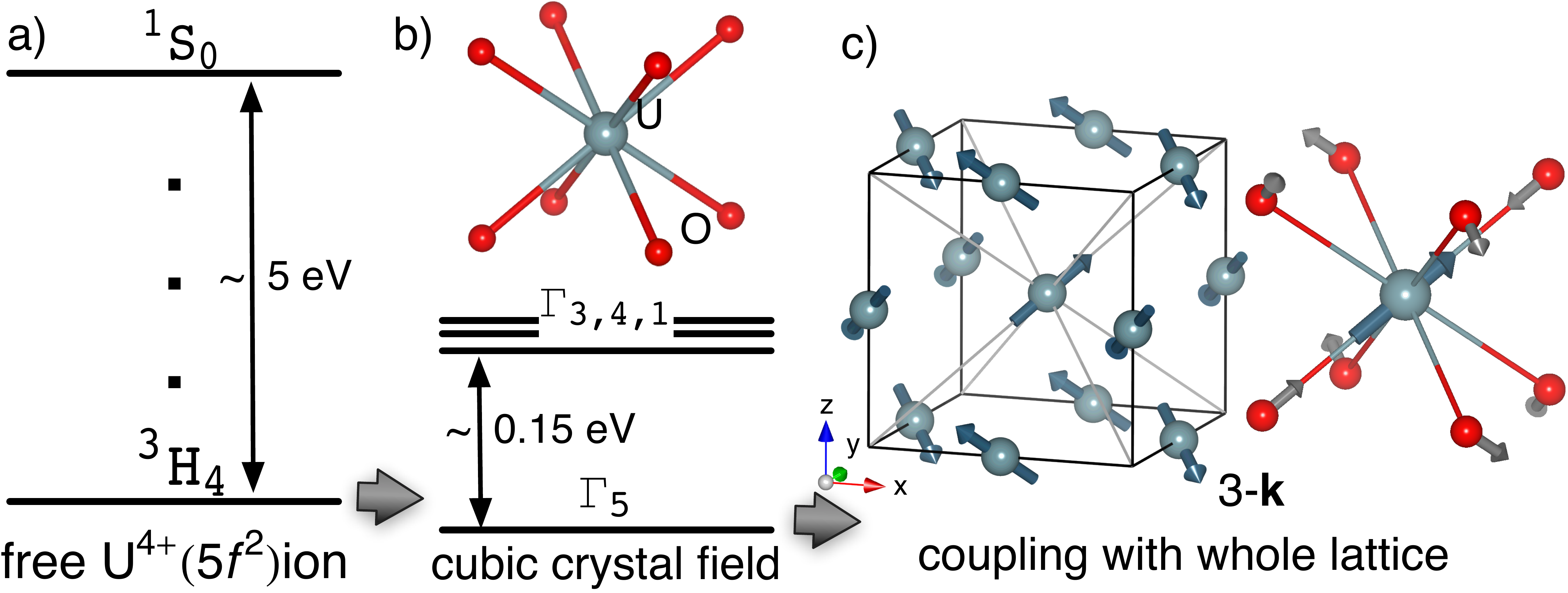}
\caption{Schematics of the $5f^{2}$ ground states and level splitting, in decreasing interaction strength, of (a) free U$^{4+}$ ion, (b) cubic CF, and (c)  ordered 3-k non-collinear magnetic structure of bulk UO$_{2}$: left, direction of magnetic moments on uranium, designated by large arrows; right, distortion of oxygen (small arrows) around a central U atom. %Going from left to right corresponds to decreasing interaction strength.
}
\label{fig:schematics}
\end{figure}

Uranium dioxide is an important and interesting material from both technological and scientific perspectives.
During the past half-century, the electronic structure of UO$_{2}$ has been thoroughly characterized by various experiments \cite{Schoenes1980PR301, Frazer1965PR1448, Willis1965PL188, Cowley1968PR464, Faber1975PRL1770, Kern1985PRB3051, Burlet1986JLM121, Osborn1988JPCSS931, Amoretti1989PRB1856, Ikushima2001PRB104404, Blackburn2005PRB184411, Wilkins2006PRB60406} (for a recent review, see Ref.~\onlinecite{Santini2009RMP807}).  
UO$_{2}$ is a semiconductor with a 2 eV band gap \cite{Schoenes1980PR301} and localized $5f^{2}$-electrons that retain strong atomic-like properties. Due to significant Coulomb interactions and spin-orbit (SO) effects, the ground state of a free U$^{4+}$ ion is the $^{3}H_{4}$ nonet [see Fig.~\ref{fig:schematics}(a)]. When the crystal-field (CF) of UO$_{2}$'s fluorite structure is considered, $^{3}H_{4}$ is split into the ground state $\Gamma_{5}$ triplet and the excited $\Gamma_{3}$ doublet, $\Gamma_{4}$ triplet, and $\Gamma_{1}$ singlet, all approximately 0.15 eV above $\Gamma_{5}$\cite{Kern1985PRB3051, Amoretti1989PRB1856}  [see Fig.~\ref{fig:schematics}(b)]. 
When cooled below $T_{N}=30.8$ K, UO$_{2}$ undergoes a first-order phase transition from a paramagnetic
%with effective magnetic moment of 3.2 $\mu_{B}$ per cation 
to a transverse type-I antiferromagnetic (AF) phase \cite{Frazer1965PR1448}, which exhibits a Jahn-Teller (JT) distortion of the oxygen cage \cite{Faber1975PRL1770}. Experimental studies now converge on the view that the non-collinear magnetic structure and the oxygen distortion are of the 3-k type \cite{Burlet1986JLM121, Amoretti1989PRB1856, Ikushima2001PRB104404, Blackburn2005PRB184411}, i.e., the moment and lattice distortion are both along the $\langle 111 \rangle$ direction [see Fig.~\ref{fig:schematics}(c)], instead of the previously proposed 1-k ($\langle 001 \rangle$) \cite{Allen1968PR530,*Allen1968PR492} and 2-k ($\langle 110 \rangle$) \cite{Faber1975PRL1770} structures.

\begin{figure}	[htbp] 
	\includegraphics[width=0.45 \linewidth]{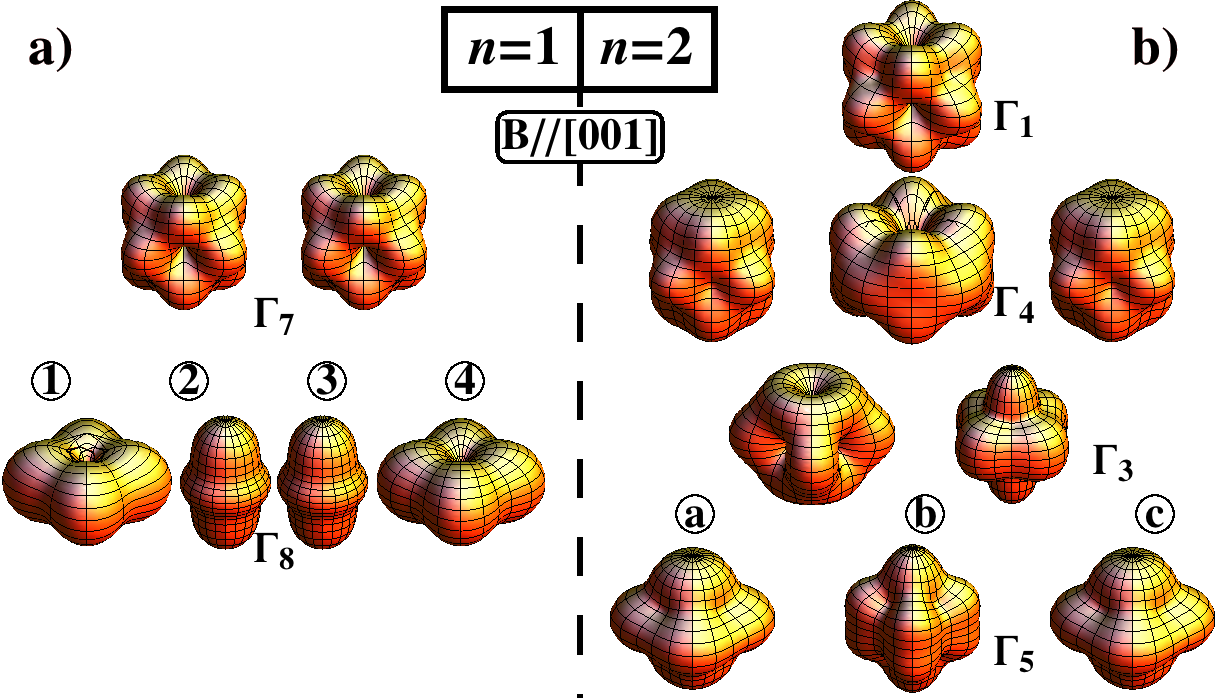} 
	\caption{Low energy $f^{n}$ eigenstates of Hamiltonian \ref{eq:CI-Hamiltonian} a) $n=1$: $\Gamma_{8}$ ground states and $\Gamma_{7}$ doublet of the $j=5/2$ sextet b) $n=2$: $\Gamma_{5}$ ground states and the excited $\Gamma_{3,4,1}$ of $^{3}H_{4}$ } \label{fig:eigenstates} 
\end{figure}

On the theory side, the CF model of Rahman and Runciman \cite{Rahman1966JPCS1833} correctly predicted the $\Gamma_{5}$ ground state of UO$_2$ (Fig.~\ref{fig:eigenstates}). Recent CF calculations have obtained quantitative agreement with experimental excitation spectra by fitting model parameters to the measured data \cite{Rahman1998PLA306}, by adding corrections to the point charge model \cite{Gajek1988JLM351}, or by extrapolating from the fitted values for other actinide dioxides \cite{Magnani2005PRB54405}. 
Models of magnetism in UO$_{2}$, pioneered by the work of Allen \cite{Allen1968PR530,*Allen1968PR492}, have explored the delicate interplay between multipolar and Jahn-Teller effects  \cite{[{For a recent review, see }]Santini2009RMP807}.

First-principles calculations have to go beyond the local-density or generalized-gradient approximations (LDA/GGA) to the density-functional theory (DFT) to correctly reproduce the insulating character of UO$_{2}$. Existence of an energy gap was demonstrated in Ref.~\onlinecite{Kudin2002PRL266402,Prodan2007PRB33101} using the hybrid functional method \cite{Becke1993JCP1372}, in Ref.~\onlinecite{Petit2010PRB45108} using the self-interaction-corrected LDA 
\cite{Perdew1981PRB5048}, and in Ref.~\onlinecite{Dudarev1997PMB613} using the DFT+$U$ method \cite{Anisimov1991PRB943}. CF splitting in actinide compounds has been computed by using constrained $f$ states without full self-consistency \cite{Colarieti-Tosti2002PRB195102} or by analyzing band positions obtained from LDA/GGA calculations \cite{Novak2007PSSB3168}. The 3-k structure of UO$_{2}$ was studied by Laskowski {\it et al.\/} using DFT+$U$ \cite{Laskowski2004PRB140408}, but their results showed anomalous dependence on the $U$ parameter and both the calculated oxygen distortions and energy differences were about an order of magnitude too large. Furthermore, the 3-k state was only stable with large $U$ values and a formulation of DFT+$U$ that is usually only applied to metals. A first-principles framework for self-consistently and accurately accounting for all the different energy scales in Fig.~\ref{fig:schematics} does not yet exist.

In this paper, we present a unified DFT-based framework for calculating the electronic spectra, magnetism and lattice distortions in UO$_{2}$. 
%Many-body electronic interactions and crystal field effects 
% edited by F.Z.
Explicit $f$-$f$ interactions and CF effects 
are treated using a model Hamiltonian with parameters derived from self-consistent DFT+$U$ calculations. The ground state wave functions that are obtained by diagonalizing this Hamiltonian are used to set up initial conditions for self-consistent DFT+$U$ calculations of magnetism and lattice relaxations. Our approach allows us to accurately reproduce all the different energy scales in Fig.~\ref{fig:schematics}, including the $\Gamma_{5}$ ground state, $\Gamma_{3,4,1}$ excited states, as well as the energetics of competing magnetic structures, including 3-k, and their associated lattice 
%relaxations
% edited by F.Z.
distortions, all within a unified self-consistent framework.

Before moving on to the details, we would like to stress that extra care should be taken in first-principles calculations of $f$ electrons. Several challenges are encountered in DFT calculations of UO$_{2}$ (and other actinide compounds in general). First, strong $f$-$f$ interactions and a weak crystal field result in an inherently complicated many-body problem. For instance, since 5$f^{2}$ electrons hybridize weakly with the O 2$p$ bands and remain well localized, their true wavefunctions are in general multi-determinantal (see below for further discussions). 
% V.O. The meaning of this remark was not quite clear. Is the above OK?
% and KS wavefunctions should bear some resemblance. 
%Secondly, degeneracy or near degeneracy calls into question the existence of a unique Hohenberg-Kohn-Sham solution. 
%We will show that although such Kohn-Sham solutions are limited to local minima of SSDs and hence not necessarily of direct physical meaning, a comprehensive study of them can help reveal the ground and excited states of many-body $f$ systems.
Second, the higly localized nature of $f$ electrons tends to magnify the inaccuracies of approximate exchange-correlation functionals. We have previously shown that the self-interaction (SI) error of $f$-electrons is highly sensitive to the occupied orbital, and its removal is non-trivial in both the DFT+$U$ and hybrid-functional methods \cite{Zhou2009PRB125127}. Therefore, an improved version of DFT+$U$ \cite{Zhou2009PRB125127} is required to remove such errors ($\sim 0.1$ eV) and access weak CF effects. 
Third, the existence of a multitude of f-states often leaves DFT+U calculations
trapped in local minima, leading to difficulties in reproducibly finding the
correct electronic ground state. As a consequence, it is not
uncommon for different authors to find inconsistent and hard-to-interpret
results with large errors ($\sim 1$ eV or even larger) even when using identical
electronic-structure methods (see Refs.~\onlinecite{Shick2001JES753, Larson2007PRB45114, Jomard2008PRB75125, Amadon2008PRB155104, Ylvisaker2009PRB35103, Dorado2009PRB235125, Zhou2009PRB125127, Meredig2010PRB195128}. 
and references therein). It is likely that this issue contributed to the failure of previous studies
to reliably examine the 3-k structure of UO$_2$. \cite{Laskowski2004PRB140408} Previously, we have shown
that the local minima issue is also present in hybrid functional calculations \cite{Zhou2009PRB125127}.
In this paper, we show that the multiple minima, corresponding to different orbital
states, contain valuable physical information about $f$-electrons that can be
used to help find the true ground state and excitation spectra.

\section{Method}
\subsection{LDA+$U$ calculations}

All DFT calculations were carried out using the VASP code \cite{Kresse1999PRB1758}, GGA-PAW potentials \cite{Blochl1994PRB17953}, a cutoff energy of 450 eV, and without any symmetry constraints to allow symmetry-breaking solutions. Crystal field calculations were performed in the primitive cell of one UO$_{2}$ formula unit with a $6\times 6 \times 6$ $k$-point grid. The lattice and ionic positions were frozen at the experimental fluorite structure for crystal field calculations. These calculations, as discussed in section\ref{sec:CF}, are ferromagnetic with one uranium ion per cell. Magnetic structures were calculated in the fcc supercell (4 formula units) using a $4\times 4 \times 4$ grid, first without and then with full relaxation. Spin-orbit coupling was self-consistently incorporated for realistic comparison with experiment. 
%Calculation initialized in specific $f^{2}$ states

To remove the orbital-dependent components of self-interaction errors (SIE) of $f$-electrons, we use a formulation of the LDA+$U$ method \cite{Zhou2009PRB125127} by modifying only the exchange term, rather than both Hartree and exchange, of the LDA:
\begin{eqnarray} 
E ^{\mathrm{LDA}+U} = E ^{\mathrm{LDA}} + E_{\mathrm{X}} - E_{\mathrm{dcX}},
% =E ^{\mathrm{LDA}} +\Delta E . 
\label{eq:newlda+u}
\end{eqnarray}
where the orbital-dependent Hartree-Fock exchange
$E_{\mathrm{X}}$ contains a term that
approximately cancels the on-site SIE in the Hartree energy of localized $f$-electrons; the
remainder of the LDA Hartree energy is exact by definition and
therefore left unmodified in our approach.
The exchange double-counting term $E_{\mathrm{dcX}}$ accounts for the
LDA exchange energy and is given by a linear
combination controlled by the $c$ parameter of the exchange double-counting in the Liechtenstein \cite{Liechtenstein1995PRB5467}  scheme
and the on-site local-spin-density (LSD) exchange, conceptually similar to hybrid functional
approaches and serves the purpose of subtracting the
orbital-dependence of the LDA exchange energy.  As a result, eq.~\ref{eq:newlda+u} is
self-interaction free to high accuracy.

\begin{figure}[htbp]
\includegraphics[width=0.8 \linewidth]{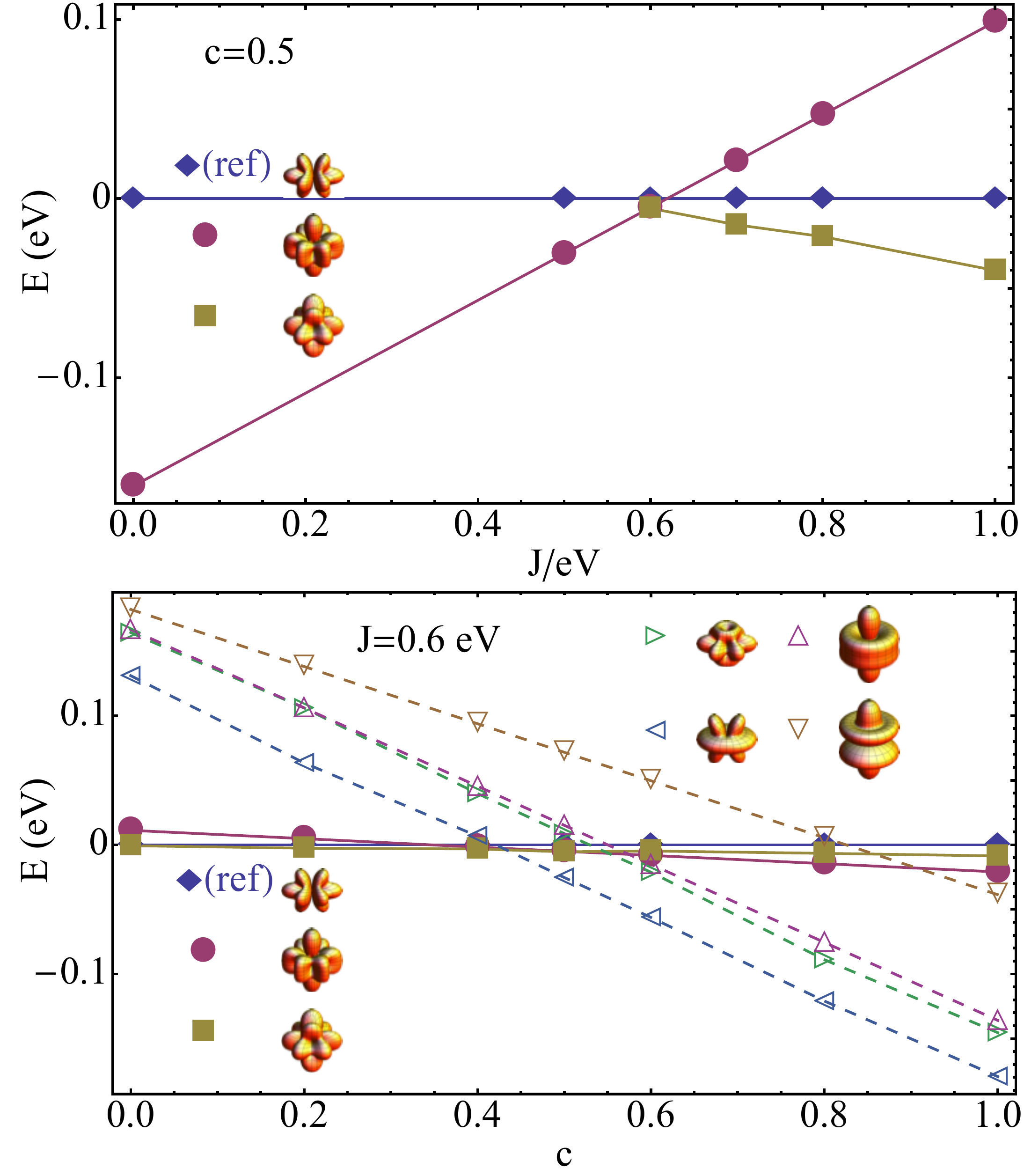}
\caption{LDA+$U$ energy of U$^{4+}$ ion as function of $J$ in different orbitals. SOC is not included.
} \label{fig:f2-degeneracy}
\end{figure}

There is only one adjustable parameter, $U$, in our approach, and the other parameters $J$ and $c$ can be determined at given $U$. As done in Ref.~\onlinecite{Zhou2009PRB125127}, we choose up to seven $f^{2}$ SSD states of the U$^{4+}$ ion that are analytically degenerate without considering spin-orbit, and calculate these states' total energy dependence on $J$ and $c$. As shown in Fig.~\ref{fig:f2-degeneracy}, optimal values of $J$=0.6 eV and $c=0.5$ are obtained at $U=$6 eV that minimize the energy difference, i.e.\ the orbital-dependent self-interaction error. These $J$ and $c$ values are used throughout the paper. We use $U$=6 eV in this paper and discuss the dependence of the results on $U$ in section \ref{sec:U}. In the rest of the paper spin-orbit is included.

%We do not need to apply the extreme LS or $jj$ coupling schemes since the DFT calculations are done self-consistently, i.e. already in the so-called intermediate coupling scheme.

\subsection{On-site model Hamiltonian for $f$}
We consider the following single-ion model for $f$-electrons: 
\begin{eqnarray}
	H_{0} &=& \sum_{i=1}^{n} ( \hat{f}_i +  \zeta \hat{\boldsymbol l}_i \cdot \hat{\boldsymbol  s}_i ) + \hat{V}_{\mathrm{ee}}, \label{eq:CI-Hamiltonian} 
\end{eqnarray}
where the summation runs over $n$ electrons for the one-body terms of cubic CF, $\hat{f}$, and SO coupling of strength $\zeta$. The electronic interaction $\hat{V}_{\mathrm{ee}}$ is parametrized by Slater's integrals $F^{k}$ ($k=0,2,4,6$) \cite{Judd1963}. The matrix elements of $\hat{f}$  between the basis states indexed by projections of orbital ($m$) and spin ($\sigma$) momenta 
are given by
\begin{eqnarray}
	\langle m \sigma |\hat{f} | m' \sigma' \rangle &=&  \delta_{\sigma \sigma'} \int \bar{Y}^{l}_{m} \left[ \frac{16\sqrt{\pi}}{3}  V_{4} (Y_{40} + \sqrt{\frac{10}{7}} \Re Y_{44}) \right. \nonumber  	\\&+& \left. 32 \sqrt{\frac{\pi}{13}} V_{6} (Y_{60} - \sqrt{14} \Re Y_{64})  \right] Y^{l}_{m'} d \Omega, 
%\langle m |\hat{f} | m' \rangle &=& -   \int \bar{Y}_{lm} \left( \lambda_{4} \alpha_{4} 
%	+ ??  \lambda_{6} \alpha_{6} \right) Y_{lm'} d \Omega, 
	\label{eq:CF-matrix-element} 
\end{eqnarray}
where $V_{4,6}$ are cubic CF parameters \cite{Newman2000} and $Y^{l}_{m}$ are complex spherical harmonics. 
%where $V_{k} (= A_{k} \langle r^{i} \rangle, \ k=4,6$)  in the conventional notation.
To study the magnetic properties, an infinitesimal magnetization field ${\boldsymbol B}$ ($\mathrm{B} \rightarrow 0$) is applied: 
\begin{eqnarray}
H' = H_{0} - \sum_{i=1}^{{n}}{\boldsymbol B}  \cdot ({g_{L} \hat{\boldsymbol l}_{i} + g_{S} \hat{{\boldsymbol s}}_{i}})  \mu_{B}/\hbar, 
%H' = H_{0} - \sum_{i=1}^{{n}}{\boldsymbol B}  \cdot ({ \hat{\boldsymbol l}_{i} + 2 \hat{{\boldsymbol s}}_{i}})  \mu_{B}/\hbar.
\label{eq:Bfield}
\end{eqnarray}
where $g_{L}=1$ and $g_{S}\approx 2$ are the orbital and spin $g$-factors, respectively. 
%Here we choose to express the CF directly with single-particle instead of $J$ eigenstates that is known as the Steven's operator equivalent method, since we work with these states in the calculations.
%Although delicate effects in the solid such as hybridization and inter-cation interactions are left out, Eq.~(\ref{eq:CI-Hamiltonian}) can fairly accurately represent on-site $f$ electron physics (see e.g.\ eq.~(104) of Ref.~\cite{Hotta2006RPP2061}).

\begin{table}[htbp]
\newlength{\graphheight} 
\graphheight 0.57cm
	\begin{ruledtabular}
		\begin{tabular}{|c|lcc|lccc|}
\multirow{2}{*}{$\boldsymbol B $} & \multicolumn{3}{c|}{$n=1$}	&\multicolumn{4}{c|}{$n=2$} \\	
          &$\Gamma_{8}$ &$\mu_{S}$ & $\mu$ & $\Gamma_{5}$ & state & $\mu_{S}$ & $\mu$ \\ \hline
		           \multirow{4}{*}{ $ [001]$ } 
& 1\includegraphics[height=\graphheight]{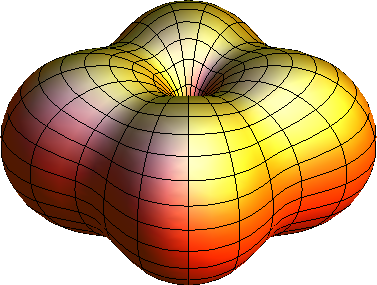}& -0.54& 1.57 & $a$ \includegraphics[height=\graphheight]{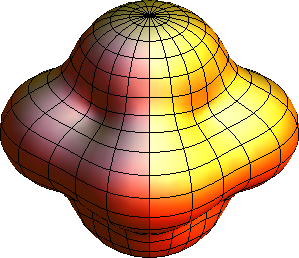} & $ 0.97 (1,2)$ 		& -0.86 & 2.06 \\
&2 \includegraphics[height=\graphheight]{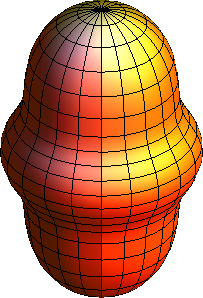}& -0.13 & 0.43 & $b$ \includegraphics[height=\graphheight]{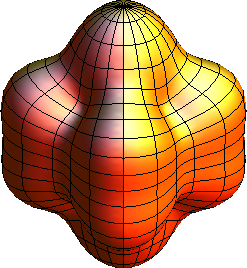} & $0.69[(1,3)$ 	& 0.00 & 0.00 \\
&3 \includegraphics[height=\graphheight]{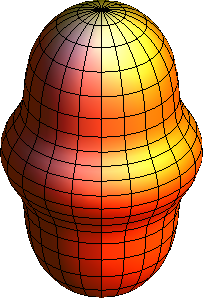}& 0.13 & -0.43 & & $+(2,4)]$ & & \\ 
&4 \includegraphics[height=\graphheight]{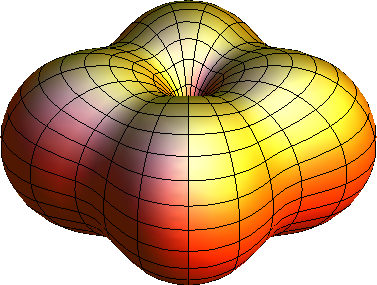}& 0.54 & -1.57 &  $c$ \includegraphics[height=\graphheight]{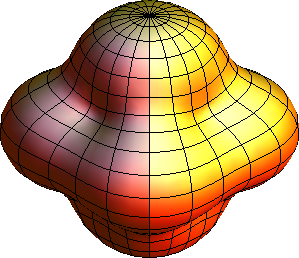} &$ 0.97 (3,4)$	 	& 0.86 & -2.06 \\  \hline
		           \multirow{4}{*}{ $ [110]$ } 
&1\includegraphics[height=\graphheight]{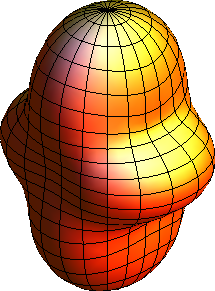} & -0.48 & 1.40 & $a $ \includegraphics[height=\graphheight]{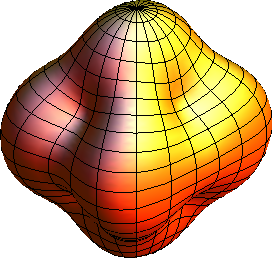} & $ 0.92 (1,2)$ 		&  -0.86 &2.06 \\
&2\includegraphics[height=\graphheight]{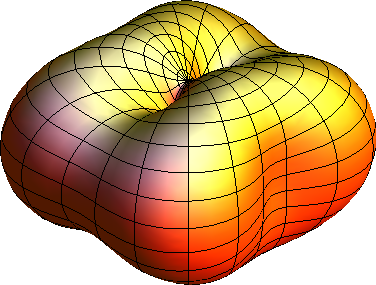}  & -0.28 & 0.83 & $b $ \includegraphics[height=\graphheight]{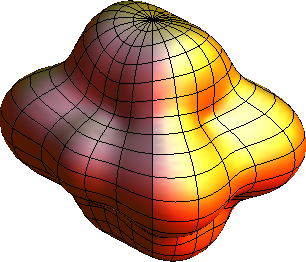} & $0.69[(1,3)$ 	&  0.00 &0.00 \\
&3\includegraphics[height=\graphheight]{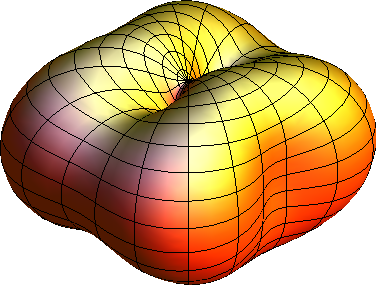}  & 0.28 & -0.83 & & $+(2,4)]$ & &   \\ 
&4\includegraphics[height=\graphheight]{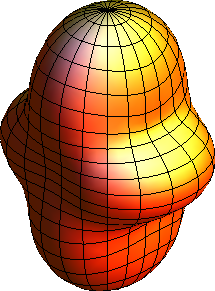}  & 0.48 & -1.40 &  $c $ \includegraphics[height=\graphheight]{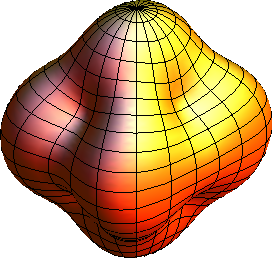} & $ 0.92 (3,4)$	&  0.86 &-2.06 \\  \hline
		           \multirow{4}{*}{ $ [111]$ } 
&1\includegraphics[height=\graphheight]{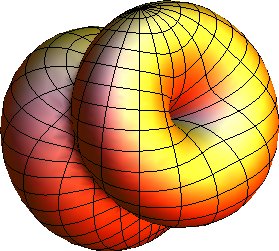}  & -0.44 & 1.28 & $a $ \includegraphics[height=\graphheight]{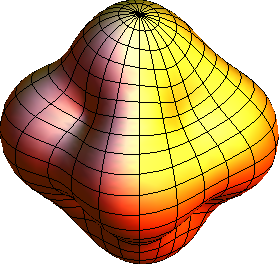} & $ 0.92 (1,2)$ 			&  -0.86 &2.06 \\
&2\includegraphics[height=\graphheight]{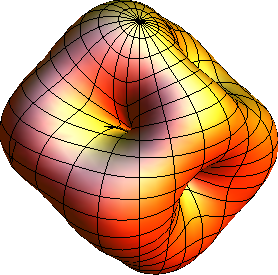}  & -0.33 & 1.00 & $b $ \includegraphics[height=\graphheight]{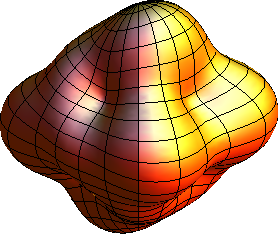} & $0.69[(1,4)$ 			&  0.00 &0.00 \\
&3\includegraphics[height=\graphheight]{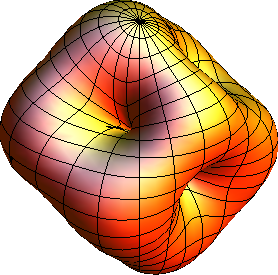}  & 0.33 & -1.00 & & $+(2,3)]$ & & \\ 
&4\includegraphics[height=\graphheight]{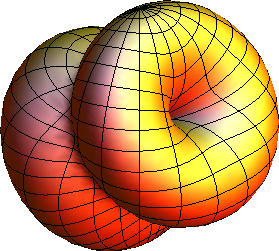}  & 0.44 & -1.28 &  $c $ \includegraphics[height=\graphheight]{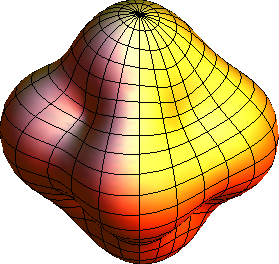} &$ 0.92 (3,4)$	&  0.86 &-2.06 \\ 
		\end{tabular}
	\end{ruledtabular}
	\caption{For different magnetization directions $\boldsymbol B $, the ground states $\Gamma_{8}$ quartet ($f^{1}$) and $\Gamma_{5}$ triplet ($f^{2}$), their spin and total magnetic moment in $\mu_{B}$, and, for $f^{2}$, the dominant determinants in the corresponding $f^{1}$ basis.
		\label{tab:fn-states}} 
\end{table}

We first discuss the general properties of solutions to Eqs.~(\ref{eq:CI-Hamiltonian})-(\ref{eq:Bfield}) using the model parameters derived from DFT+$U$ calculations (which will be discussed in detail in section \ref{sec:CF}). For $n=1$, 14 eigenstates are obtained, the lowest being the $\Gamma_8$ quartet (Fig.~\ref{fig:eigenstates} and Table \ref{tab:fn-states}). For $n>1$, the Hamiltonian in Eq.~(\ref{eq:CI-Hamiltonian}) can be diagonalized via configuration interaction of $C^{14}_n$ $f^n$ single Slater determinants (SSD) based on the $f^1$ eigenstates. The $\Gamma_{5}$ ground states of $f^{2}$ are shown in Table \ref{tab:fn-states}, together with their dominant determinants, designated as $(i,j)$ using the indices of $f^1$ states in the left column of Table \ref{tab:fn-states}. For magnetic moment along each of the $[001]$, $[110]$, and $[111]$ directions, the $\Gamma_{8}$ quartet of $f^{1}$ includes states 1,4 (2,3) with larger (smaller) spin and orbital magnetic momenta, while the $\Gamma_{5}$ triplet of $f^{2}$ consists of states $a$ and $c$ with $|\mu|=2.06$ $\mu_{B}$ and one dominant determinant [$(1,2)$ or $(3,4)$], as well as a non-magnetic state $b$ dominated by two determinants (right column of Table~\ref{tab:fn-states}). Note that the observed moment of the ordered state is smaller at 1.75 $\mu_{B}$ \cite{Faber1975PRL1770}.
The moment $\mu=2.06 \mu_{B}$ of the $\Gamma_{5}(a,c)$ states is slightly larger than the saturated $2 \mu_{B}$ characteristic of the $^{3}H_{4}$ multiplet because exchange and SOC interactions are all of comparable strength and other multiplets slightly mix into the ground state and increase the effective moment \cite{Rahman1966JPCS1833}. In general, all the $f^{2}$ eigenstates, including $\Gamma_{3,4,1}$ (Fig.~\ref{fig:eigenstates}) are composed of multiple determinants.

\subsection{Model parameters from LDA+$U$} \label{sec:fitting}
The parameters for the model Hamiltonian in Eq.~(\ref{eq:CI-Hamiltonian}) are obtained by analyzing the total energies and $f$ wavefunctions calculated with LDA+$U$. In this procedure, many self-consistent LDA+$U$ calculations are first carried out, yielding solutions that are in general local minima rather than the global minimum of UO$_{2}$. Next, we extract the $f^{2}$ Kohn-Sham wavefunction $|\Psi_{f} \rangle$, SSD by construction, from each solution, and compute the expected energy  according to Eq.~(\ref{eq:CI-Hamiltonian}) 
\begin{eqnarray}
\langle \Psi_{f} | H_{0} |\Psi_{f} \rangle = x_{1} V_{4} + x_{2} V_{6} + x_{3} \zeta + x_{4} F^{2} + U, \label{eq:expected-E}
\end{eqnarray}
where $x_{i}$'s represent the solution-dependent coefficient associated with model parameters. Since the $F^{k}$ ($k=2,4,6$) contributions of $\hat{V}_{\mathrm{ee}}$ are heavily correlated \cite{Carnall1992JCP8713}, the following approximation\cite{Berry1988CP105} has been adopted in Eq.~(\ref{eq:expected-E}):  
\begin{eqnarray}
F^2&=&F^4/0.668=F^6/0.494,  %\\
%&=& 6435J/(286+195 \times 0.668+250 \times 0.494),
\label{eq:Fk-to-J}
\end{eqnarray}
eliminating model parameters  $F^{4}$ and $F^{6}$. Finally, expectation values of $H$ (Eq.~\ref{eq:expected-E}) of the obtained solutions are fitted to the corresponding DFT+$U$ total energies, yielding self-consistent {ab initio} values of $F^{2}$, $\zeta$ and the crystal field parameters $V_{4}$ and $V_{6}$. We use the simple least-square method to perform the linear fitting. Here $U$ in eq.~(\ref{eq:expected-E}) can be regarded as a constant in the fitting and bears no direct physical meaning.

\section{Results and discussions}

\subsection{Crystal field ground states and excitations} \label{sec:CF}
\begin{figure}[htbp]
\includegraphics[width=0.85 \linewidth]{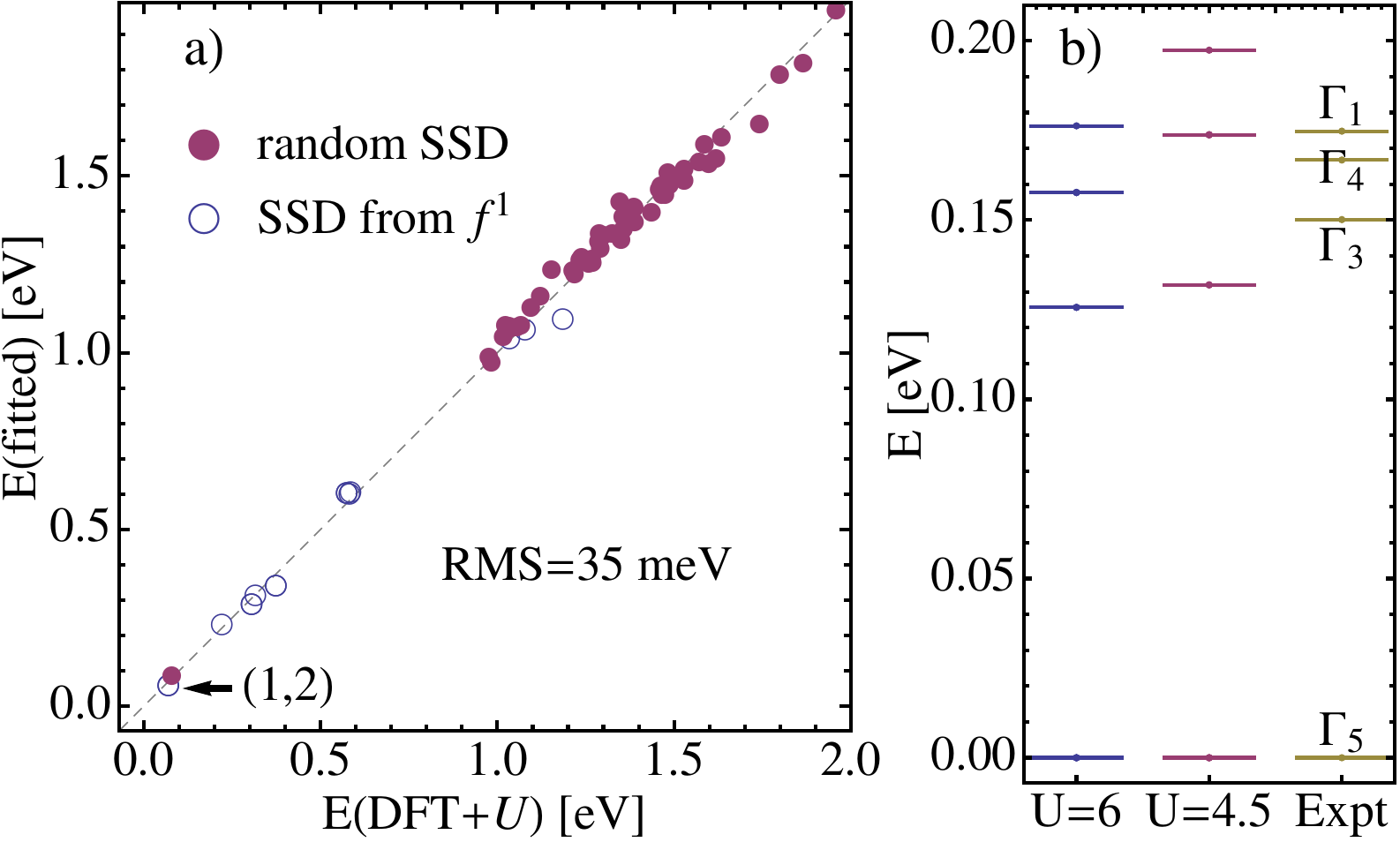}
\caption{a) Fitting of DFT+$U$ \cite{Zhou2009PRB125127} energy to Eq.~\ref{eq:CI-Hamiltonian} for 50 runs with random initial states (filled circles) and 15 states with initial states constructed from $f^{1}$ solutions (open circles). b)   Predicted $f^{2}$ CF levels $\Gamma_{5,3,4,1}$ compared to measured CF splitting \cite{Amoretti1989PRB1856}.}
\label{fig:fitting-prediction}
\end{figure}

We carried out a series of 50 different self-consistent calculations with randomly initialized $f^{2}$ states. Due to the existence of multiple local minima in DFT+$U$, these calculations resulted in a range of energies spread over almost 2 eV [filled circles in Fig.~\ref{fig:fitting-prediction}(a)]. It is seen that random wavefunction initialization has generated only one low-energy solution, while the remaining runs were trapped in metastable high-energy states.

Obtained from the fitting procedure outlined in section \ref{sec:fitting}, model parameters are applied in Eq.~(\ref{eq:CI-Hamiltonian}) to construct $f^{1}$ eigenstates and subsequently determine $f^{2}$ states by direct diagonalization within the subspace of SSDs formed from $f^{1}$ eigenstates. 
To further improve the quality of our fit and provide data points in the low-energy region that was poorly represented in the randomly initialized sample [filled circles in Fig.~\ref{fig:fitting-prediction}(a)], we self-consistently calculate the DFT$+U$ energies of additional 15 two-electron SSD states that involve the 6 low energy $f^{1}$ orbitals ($\Gamma_{8}$ and $\Gamma_{7}$) for ${\boldsymbol B}//[001]$; these points are shown as open circles in Fig.~\ref{fig:fitting-prediction}a. The $(1,2)$ and $(3,4)$ states, which dominate the $\Gamma_{5a}$ and  $\Gamma_{5c}$ ground states, are also found to have the lowest energies in self-consistent DFT$+U$ calculations, demonstrating that our method can reliably locate the electronic ground state. 
The other ground state in Table~\ref{tab:fn-states}, $\Gamma_{5b}$, has two dominant determinants and is not directly accessible in DFT+$U$. 
Therefore, the data flow between the model Hamiltonian and DFT$+U$ calculations is bi-directional: DFT$+U$ provides model parameters, while the model guides the DFT$+U$ to the ground state and gives access to multi-determinant states.

\begin{table}[htbp] 
	\begin{ruledtabular}
		\begin{tabular}{|l|cccccc|} 
		&$F^{2}$	&$F^{4}$	&$F^{6}$	& $\zeta$	& $V_{4}$	& $V_{6}$	\\
$U=6$ eV 	 & 5.649 & (3.773) & (2.790) & 0.230 & -0.093 & 0.0157 \\
$U=4.5$ eV	& 5.495 & (3.670) & (2.714) & 0.209 & -0.106 & 0.0163\\
U$^{4+}$ ion\cite{VanDeurzen1984JOSAB45} & 6.439 & 5.295 & 3.440 & 0.244 & & \\
Ref.~\cite{Rahman1998PLA306} & \multicolumn{4}{c}{using ref.~\cite{VanDeurzen1984JOSAB45}} & -0.112& 0.024 \\
Ref.~\cite{Amoretti1989PRB1856}  &  &&&	& -0.123	& 0.0265 \\
Ref.~\cite{Magnani2005PRB54405}  &  	& 	& 	&  	& -0.155	& 0.0333 \\
		\end{tabular}
	\end{ruledtabular}
	\caption{Fitted parameters (in eV) using Eqs.~(\ref{eq:CI-Hamiltonian}),  (\ref{eq:CF-matrix-element}), and (\ref{eq:Fk-to-J}), compared with prior studies.
	\label{tab:parameters}} 
\end{table}

The final fitted parameters are shown in Table \ref{tab:parameters}. Compared to the values obtained by fitting the spectra of free ions \cite{VanDeurzen1984JOSAB45}, the ionic parameters $F^{k}$ and $\zeta$ in the UO$_{2}$ solid are somewhat suppressed due to hybridization and screening effects. The calculated cubic CF parameters $V_{4}$ and  $V_{6}$ are slightly smaller than those fitted to experimental data or extrapolated from other actinide oxides \cite{Rahman1998PLA306, Amoretti1989PRB1856, Magnani2005PRB54405}. The Hamiltonian in Eq.~(\ref{eq:CI-Hamiltonian}) can now be diagonalized.
The predicted energies of the three lowest excited CF levels $\Gamma_{3,4,1}$ are in reasonable agreement with experiment \cite{Amoretti1989PRB1856} with errors of approximately 10-20 meV [Fig.~\ref{fig:fitting-prediction}(b)]. A notable deviation is overestimation of the splitting between these levels.

Finally, we note that the input $J=0.6$ eV used in our DFT+$U$ calculation differs from the fitted value of $J' =(286 F^{2}+195 F^{4}+250 F^{6})/6435$ in Table \ref{tab:parameters}. This is because the role of the former is to minimize the SI error in DFT$+U$, while the latter represents on-site exchange, and some difference between them is expected when used with an approximate xc functional. A perfect $xc$ functional would make the input $U$ or $J$ unnecessary and predict physically meaningful output $J'$ or $F^{k}$.

\subsection{Dependence on input $U$} \label{sec:U}

To illustrate the effect of the only adjustable variable in our approach, $U$, the same calculations were repeated using $U=4.5$ eV. As shown in Table~\ref{tab:parameters} and Fig.~\ref{fig:fitting-prediction}(b), the results change only slightly and remain in good agreement with experiment. Note that when $U$, which controls the degree of electron localization, is decreased, the ionic parameters $F^{k}$ and $\zeta$ also decrease, i.e.\ away from the free ion values,
while the CF parameters increase, suggesting that the $f$-electrons become more delocalized. Such a picture of opposite influence of electron localization on free ion and CF parameters is consistent with the observed trend that increase of the CF interaction results in a reduction in the free-ion parameters for the same ion in different chemical environment \cite{Liu2005JSSC489}.

\subsection{Magnetic properties}
Finally, we discuss the effects of inter-ionic interactions and magnetic properties of UO$_2$. Various magnetic structures within a cubic supercell of four formula units were calculated, first without and then with lattice relaxation. The previous approximation of representing the $\Gamma_{5}$ ground state with the $(1,2),(3,4)$ SSDs was adopted. Table~\ref{tab:magnetic} shows the energies $E_{s}$ (static lattice), $E_{r}$ (after full relaxation), and the total magnetic moments $\mu$ assuming ferromagnetic (FM) and type-A anti-ferromagnetic (AAF) configurations along the $[001]$ (reference), $[110]$, and $[111]$ directions.  In all cases, the calculated $\mu$ of  $\sim$ 2.1--2.2 $\mu_{B}$ is close to the saturated value 2.06 $\mu_{B}$ in Table \ref{tab:fn-states} and larger than the measured 1.74 $\mu_{B}$ \cite{Frazer1965PR1448}. The reduction of the ordered moment is a topic of considerable interest, but the mechanisms, such as the dynamical Jahn-Teller effect \cite{Ippolito2005PRB64419}, are beyond the scope of this work. 
Given that the $(1,2)/(3,4)$ states are the CF ground states within our computational approach and the calculated moment is not too far from 1.74 $\mu_{B}$, we continue to use these settings for non-collinear magnetic calculations.

Table~\ref{tab:magnetic} shows that for each magnetization direction, the AAF configuration is always lower than the FM configuration, in agreement with experiment. The energy differences $E_{s}(\mathrm{FM})-E_{s}(\mathrm {AAF})$ are in the 4-6 meV range, suggesting that multipolar interactions and anisotropy are weak.
These weak, mostly isotropic interactions underlie the success of the simple fitting procedure of Section \ref{sec:fitting}  (root-mean-square error = 35 meV) in FM configurations of 15 $\langle 001 \rangle$ solutions and 50 solutions with random magnetic moment directions. We also find that the different magnetization directions differ in energy by less than $9$ meV, suggesting that our procedure for removing the orbital-depedent SI error is highly accurate. Indeed, the $f^{2}$ wavefunctions differ considerably for the three principle directions (see Table \ref{tab:fn-states}), which would result in SI errors of $0.1 \sim 0.2$ eV using the unmodified DFT$+U$ approach. Nevertheless, we take additional care to remove any remaining SI errors, however small they they appear to be, by subtracting a reference energy $E_{d}$ for each principle magnetization direction $d=\langle 001 \rangle$, $\langle 110 \rangle$, or $\langle 111 \rangle$: $E_{d}=[E_{s}(\mathrm{FM})+E_{s}(\mathrm {AAF})]/2.$ With this correction, the AAF configurations in the three directions, as well as the 3-k structure (magnetic moment along $\langle 111 \rangle$, see Fig.~\ref{fig:schematics}c), are essentially degenerate before lattice relaxation.

The magnetic transition temperature of antiferromagnetic UO$_{2}$ is estimated with a classical Heisenberg model on an fcc lattice:
\begin{equation}
\mathcal{H}= - J_{H} \sum_{\langle i j \rangle} \vec{s}_{i} \cdot \vec{s}_{j},
\end{equation}
where the summation is over all nearest neighbor sites $\langle i j \rangle$ with unit spin $\vec{s}$. The FM/AAF energy difference per UO$_{2}$ is $\Delta E= 6 J_{H} - (-2 J_{H}) = 8 J_{H}$.  As shown in Table~\ref{tab:magnetic}, $\Delta E \approx 6$ meV, corresponding to $T_{N} = 3.18 J_{H}/k_{B}$ \cite{McKenzie1982JPA3899}, or about 28 K, in excellent agreement with the experimental value 30.8 K.\cite{Frazer1965PR1448}

After relaxation without symmetry constraints, we obtain the energies listed in the right column of Table~\ref{tab:magnetic}. The computed moments $\mu$ become slightly larger than the static values. The relaxation energy $E_{r}$ and the corrected $E_{r}-E_{d}$ are large ($>50$ meV) due to differences between the static (fixed to experimental $a=5.47$~{\AA}) and relaxed lattice parameters. The energy differences between competing magnetic configurations increases to $\sim 10$ meV; these values are consistent with the N{\'{e}}el temperature of $T_{N}=30.8$ K \cite{Frazer1965PR1448}. The relaxed structures with $\mu // \langle 111 \rangle$ are clearly more stable than $\langle 001 \rangle$ and $\langle 110 \rangle$. We have enumerated all AF $\langle 111 \rangle$ structures within the fcc unit cell and found that the 3-k structure indeed has the lowest energy. The associated oxygen lattice distortion (amplitude 0.024 \AA) is also of the 3-k type, though slightly larger than the measured 0.014 \AA \cite{Faber1975PRL1770}.

%JT cooperative effect and exchange interactions are antagonists since the JT effect tends to stabilize a state with the quenched orbital momentum on each metal ion while the exchange interactions acts in a just opposite direction, i.e., it tends to stabilize a state with the maximum orbital momentum (which is strongly coupled with the spin via the SO interaction).

\begin{table}[htbp] 
	\begin{ruledtabular}
		\begin{tabular}{|l|ccc|ccc|} 
	&\multicolumn{3}{c|}{static}	&\multicolumn{3}{c|}{relaxed} \\
config. 	& $E_{s}$ & $E_{s}- E_{d} $ & $|\mu|/\mu_{B}$ & $E_{r}$ & $E_{r}- E_{d} $ & $|\mu|/\mu_{B}$ \\ \hline
$[001]$ FM 	& 0 (ref)	& 3.3  & 2.11  &  -57.3 &  -54.0 	& 2.20   \\
$[001]$ AAF 	& -6.5  	& -3.3  &2.11   &-61.9   &  -58.6	& 2.35   \\ \hline
$[110]$ FM 	& -8.2 	& 2.1  &2.15   &-65.5   & -55.2  	& 2.33   \\
$[110]$ AAF 	& -12.4	& -2.1  &2.15  &-71.8	   & -61.5	& 2.22   \\ \hline
$[111]$ FM    	& -8.5	& 2.4	   &2.22  &-70.3	   & -59.5	& 2.22   \\
$[111]$ AAF 	& -13.2	& -2.4  &2.21  &-76.3   & -65.5	& 2.27   \\
3-k  & -13.8  & -3.0  & 2.21  & -81.1   & {\bf{-70.3}}  & 2.39  
		\end{tabular}
	\end{ruledtabular}
	\caption{Energy (in meV per UO$_{2}$) for different magnetic structures, without and with ionic relaxation. 
	%Longitudinal (L) and transverse (T) TA, TB models for 3-k: Fig.~9 of Ref.~\cite{Santini2009RMP807}
	\label{tab:magnetic}} 
\end{table}

\section{Conclusions}
In summary, we have studied the electronic structure of UO$_2$ using
an aspherical-self-interaction free DFT+$U$ method coupled with a model Hamiltonian.
The $\Gamma_{5}$ crystal field ground states, as well as the CF excitation energies are reproduced in good agreement with
experiment. Various magnetic structures are investigated with
careful initialization of the orbital and magnetic states. The inter-ionic
interactions are found to be weak and largely isotropic.
When self-interaction errors are accounted for, the 3-k
structure is essentially degenerate with other
antiferromagnetic configurations and becomes the ground state only
when lattice relaxations are considered. Our work demonstrates the
usefulness of electronic structure calculations for $f$-compounds with proper treatment of self-interaction errors and multiple
self-consistent local minima corresponding to different orbital states;
this approach can be readily applied to defect supercells and other $f$-compounds.

\begin{acknowledgments}
This work was supported by the U.S. Department of Energy, Nuclear Energy Research Initiative Consortium (NERI-C) under grant No.\ DE-FG07-07ID14893, and used resources of the National Energy Research Scientific Computing Center (NERSC).
\end{acknowledgments}

%\bibliography{../../../Documents/manuscript/newbib,../../../Documents/manuscript/other}

\begin{thebibliography}{48}%
\makeatletter
\providecommand \@ifxundefined [1]{%
 \@ifx{#1\undefined}
}%
\providecommand \@ifnum [1]{%
 \ifnum #1\expandafter \@firstoftwo
 \else \expandafter \@secondoftwo
 \fi
}%
\providecommand \@ifx [1]{%
 \ifx #1\expandafter \@firstoftwo
 \else \expandafter \@secondoftwo
 \fi
}%
\providecommand \natexlab [1]{#1}%
\providecommand \enquote  [1]{``#1''}%
\providecommand \bibnamefont  [1]{#1}%
\providecommand \bibfnamefont [1]{#1}%
\providecommand \citenamefont [1]{#1}%
\providecommand \href@noop [0]{\@secondoftwo}%
\providecommand \href [0]{\begingroup \@sanitize@url \@href}%
\providecommand \@href[1]{\@@startlink{#1}\@@href}%
\providecommand \@@href[1]{\endgroup#1\@@endlink}%
\providecommand \@sanitize@url [0]{\catcode `\\12\catcode `\$12\catcode
  `\&12\catcode `\#12\catcode `\^12\catcode `\_12\catcode `\%12\relax}%
\providecommand \@@startlink[1]{}%
\providecommand \@@endlink[0]{}%
\providecommand \url  [0]{\begingroup\@sanitize@url \@url }%
\providecommand \@url [1]{\endgroup\@href {#1}{\urlprefix }}%
\providecommand \urlprefix  [0]{URL }%
\providecommand \Eprint [0]{\href }%
\providecommand \doibase [0]{http://dx.doi.org/}%
\providecommand \selectlanguage [0]{\@gobble}%
\providecommand \bibinfo  [0]{\@secondoftwo}%
\providecommand \bibfield  [0]{\@secondoftwo}%
\providecommand \translation [1]{[#1]}%
\providecommand \BibitemOpen [0]{}%
\providecommand \bibitemStop [0]{}%
\providecommand \bibitemNoStop [0]{.\EOS\space}%
\providecommand \EOS [0]{\spacefactor3000\relax}%
\providecommand \BibitemShut  [1]{\csname bibitem#1\endcsname}%
\let\auto@bib@innerbib\@empty
%</preamble>
\bibitem [{\citenamefont {Schoenes}(1980)}]{Schoenes1980PR301}%
  \BibitemOpen
  \bibfield  {author} {\bibinfo {author} {\bibfnamefont {J.}~\bibnamefont
  {Schoenes}},\ }\href@noop {} {\bibfield  {journal} {\bibinfo  {journal}
  {Phys. Rep.}\ }\textbf {\bibinfo {volume} {63}},\ \bibinfo {pages} {301}
  (\bibinfo {year} {1980})}\BibitemShut {NoStop}%
\bibitem [{\citenamefont {Frazer}\ \emph {et~al.}(1965)\citenamefont {Frazer},
  \citenamefont {Shirane}, \citenamefont {Cox},\ and\ \citenamefont
  {Olsen}}]{Frazer1965PR1448}%
  \BibitemOpen
  \bibfield  {author} {\bibinfo {author} {\bibfnamefont {B.~C.}\ \bibnamefont
  {Frazer}}, \bibinfo {author} {\bibfnamefont {G.}~\bibnamefont {Shirane}},
  \bibinfo {author} {\bibfnamefont {D.~E.}\ \bibnamefont {Cox}}, \ and\
  \bibinfo {author} {\bibfnamefont {C.~E.}\ \bibnamefont {Olsen}},\ }\href@noop
  {} {\bibfield  {journal} {\bibinfo  {journal} {Phys. Rev.}\ }\textbf
  {\bibinfo {volume} {140}},\ \bibinfo {pages} {A1448} (\bibinfo {year}
  {1965})}\BibitemShut {NoStop}%
\bibitem [{\citenamefont {WILLIS}\ and\ \citenamefont
  {Taylor}(1965)}]{Willis1965PL188}%
  \BibitemOpen
  \bibfield  {author} {\bibinfo {author} {\bibfnamefont {B.~T.~M.}\
  \bibnamefont {WILLIS}}\ and\ \bibinfo {author} {\bibfnamefont {R.~I.}\
  \bibnamefont {Taylor}},\ }\href@noop {} {\bibfield  {journal} {\bibinfo
  {journal} {Phys. Lett.}\ }\textbf {\bibinfo {volume} {17}},\ \bibinfo {pages}
  {188} (\bibinfo {year} {1965})}\BibitemShut {NoStop}%
\bibitem [{\citenamefont {COWLEY}\ and\ \citenamefont
  {Dolling}(1968)}]{Cowley1968PR464}%
  \BibitemOpen
  \bibfield  {author} {\bibinfo {author} {\bibfnamefont {R.~A.}\ \bibnamefont
  {COWLEY}}\ and\ \bibinfo {author} {\bibfnamefont {G.}~\bibnamefont
  {Dolling}},\ }\href@noop {} {\bibfield  {journal} {\bibinfo  {journal} {Phys.
  Rev.}\ }\textbf {\bibinfo {volume} {167}},\ \bibinfo {pages} {464} (\bibinfo
  {year} {1968})}\BibitemShut {NoStop}%
\bibitem [{\citenamefont {Faber}\ \emph {et~al.}(1975)\citenamefont {Faber},
  \citenamefont {Lander},\ and\ \citenamefont {Cooper}}]{Faber1975PRL1770}%
  \BibitemOpen
  \bibfield  {author} {\bibinfo {author} {\bibfnamefont {J.}~\bibnamefont
  {Faber}}, \bibinfo {author} {\bibfnamefont {G.~H.}\ \bibnamefont {Lander}}, \
  and\ \bibinfo {author} {\bibfnamefont {B.~R.}\ \bibnamefont {Cooper}},\
  }\href@noop {} {\bibfield  {journal} {\bibinfo  {journal} {Phys. Rev. Lett.}\
  }\textbf {\bibinfo {volume} {35}},\ \bibinfo {pages} {1770} (\bibinfo {year}
  {1975})}\BibitemShut {NoStop}%
\bibitem [{\citenamefont {Kern}\ \emph {et~al.}(1985)\citenamefont {Kern},
  \citenamefont {Loong},\ and\ \citenamefont {Lander}}]{Kern1985PRB3051}%
  \BibitemOpen
  \bibfield  {author} {\bibinfo {author} {\bibfnamefont {S.}~\bibnamefont
  {Kern}}, \bibinfo {author} {\bibfnamefont {C.~K.}\ \bibnamefont {Loong}}, \
  and\ \bibinfo {author} {\bibfnamefont {G.~H.}\ \bibnamefont {Lander}},\
  }\href@noop {} {\bibfield  {journal} {\bibinfo  {journal} {Phys. Rev. B}\
  }\textbf {\bibinfo {volume} {32}},\ \bibinfo {pages} {3051} (\bibinfo {year}
  {1985})}\BibitemShut {NoStop}%
\bibitem [{\citenamefont {Burlet}\ \emph {et~al.}(1986)\citenamefont {Burlet},
  \citenamefont {ROSSATMIGNOD}, \citenamefont {QUEZEL}, \citenamefont {VOGT},
  \citenamefont {SPIRLET},\ and\ \citenamefont {Rebizant}}]{Burlet1986JLM121}%
  \BibitemOpen
  \bibfield  {author} {\bibinfo {author} {\bibfnamefont {P.}~\bibnamefont
  {Burlet}}, \bibinfo {author} {\bibfnamefont {J.}~\bibnamefont
  {ROSSATMIGNOD}}, \bibinfo {author} {\bibfnamefont {S.}~\bibnamefont
  {QUEZEL}}, \bibinfo {author} {\bibfnamefont {O.}~\bibnamefont {VOGT}},
  \bibinfo {author} {\bibfnamefont {J.~C.}\ \bibnamefont {SPIRLET}}, \ and\
  \bibinfo {author} {\bibfnamefont {J.}~\bibnamefont {Rebizant}},\ }\href@noop
  {} {\bibfield  {journal} {\bibinfo  {journal} {J. Less-Common. Met.}\
  }\textbf {\bibinfo {volume} {121}},\ \bibinfo {pages} {121} (\bibinfo {year}
  {1986})}\BibitemShut {NoStop}%
\bibitem [{\citenamefont {Osborn}\ \emph {et~al.}(1988)\citenamefont {Osborn},
  \citenamefont {Taylor}, \citenamefont {Bowden}, \citenamefont {Hackett},
  \citenamefont {Hayes}, \citenamefont {Hutchings}, \citenamefont {Amoretti},
  \citenamefont {Caciuffo}, \citenamefont {Blaise},\ and\ \citenamefont
  {Fournier}}]{Osborn1988JPCSS931}%
  \BibitemOpen
  \bibfield  {author} {\bibinfo {author} {\bibfnamefont {R.}~\bibnamefont
  {Osborn}}, \bibinfo {author} {\bibfnamefont {A.~D.}\ \bibnamefont {Taylor}},
  \bibinfo {author} {\bibfnamefont {Z.~A.}\ \bibnamefont {Bowden}}, \bibinfo
  {author} {\bibfnamefont {M.~A.}\ \bibnamefont {Hackett}}, \bibinfo {author}
  {\bibfnamefont {W.}~\bibnamefont {Hayes}}, \bibinfo {author} {\bibfnamefont
  {M.~T.}\ \bibnamefont {Hutchings}}, \bibinfo {author} {\bibfnamefont
  {G.}~\bibnamefont {Amoretti}}, \bibinfo {author} {\bibfnamefont
  {R.}~\bibnamefont {Caciuffo}}, \bibinfo {author} {\bibfnamefont
  {A.}~\bibnamefont {Blaise}}, \ and\ \bibinfo {author} {\bibfnamefont {J.~M.}\
  \bibnamefont {Fournier}},\ }\href@noop {} {\bibfield  {journal} {\bibinfo
  {journal} {J Phys C Solid State}\ }\textbf {\bibinfo {volume} {21}},\
  \bibinfo {pages} {L931} (\bibinfo {year} {1988})}\BibitemShut {NoStop}%
\bibitem [{\citenamefont {Amoretti}\ \emph {et~al.}(1989)\citenamefont
  {Amoretti}, \citenamefont {Blaise}, \citenamefont {Caciuffo}, \citenamefont
  {Fournier}, \citenamefont {Hutchings}, \citenamefont {Osborn},\ and\
  \citenamefont {Taylor}}]{Amoretti1989PRB1856}%
  \BibitemOpen
  \bibfield  {author} {\bibinfo {author} {\bibfnamefont {G.}~\bibnamefont
  {Amoretti}}, \bibinfo {author} {\bibfnamefont {A.}~\bibnamefont {Blaise}},
  \bibinfo {author} {\bibfnamefont {R.}~\bibnamefont {Caciuffo}}, \bibinfo
  {author} {\bibfnamefont {J.~M.}\ \bibnamefont {Fournier}}, \bibinfo {author}
  {\bibfnamefont {M.~T.}\ \bibnamefont {Hutchings}}, \bibinfo {author}
  {\bibfnamefont {R.}~\bibnamefont {Osborn}}, \ and\ \bibinfo {author}
  {\bibfnamefont {A.~D.}\ \bibnamefont {Taylor}},\ }\href@noop {} {\bibfield
  {journal} {\bibinfo  {journal} {Phys. Rev. B}\ }\textbf {\bibinfo {volume}
  {40}},\ \bibinfo {pages} {1856} (\bibinfo {year} {1989})}\BibitemShut
  {NoStop}%
\bibitem [{\citenamefont {Ikushima}\ \emph {et~al.}(2001)\citenamefont
  {Ikushima}, \citenamefont {Tsutsui}, \citenamefont {Haga}, \citenamefont
  {Yasuoka}, \citenamefont {Walstedt}, \citenamefont {Masaki}, \citenamefont
  {Nakamura}, \citenamefont {Nasu},\ and\ \citenamefont
  {Onuki}}]{Ikushima2001PRB104404}%
  \BibitemOpen
  \bibfield  {author} {\bibinfo {author} {\bibfnamefont {K.}~\bibnamefont
  {Ikushima}}, \bibinfo {author} {\bibfnamefont {S.}~\bibnamefont {Tsutsui}},
  \bibinfo {author} {\bibfnamefont {Y.}~\bibnamefont {Haga}}, \bibinfo {author}
  {\bibfnamefont {H.}~\bibnamefont {Yasuoka}}, \bibinfo {author} {\bibfnamefont
  {R.~E.}\ \bibnamefont {Walstedt}}, \bibinfo {author} {\bibfnamefont {N.~M.}\
  \bibnamefont {Masaki}}, \bibinfo {author} {\bibfnamefont {A.}~\bibnamefont
  {Nakamura}}, \bibinfo {author} {\bibfnamefont {S.}~\bibnamefont {Nasu}}, \
  and\ \bibinfo {author} {\bibfnamefont {K.}~\bibnamefont {Onuki}},\
  }\href@noop {} {\bibfield  {journal} {\bibinfo  {journal} {Phys. Rev. B}\
  }\textbf {\bibinfo {volume} {63}},\ \bibinfo {pages} {104404} (\bibinfo
  {year} {2001})}\BibitemShut {NoStop}%
\bibitem [{\citenamefont {Blackburn}\ \emph {et~al.}(2005)\citenamefont
  {Blackburn}, \citenamefont {Caciuffo}, \citenamefont {Magnani}, \citenamefont
  {Santini}, \citenamefont {Brown}, \citenamefont {Enderle},\ and\
  \citenamefont {Lander}}]{Blackburn2005PRB184411}%
  \BibitemOpen
  \bibfield  {author} {\bibinfo {author} {\bibfnamefont {E.}~\bibnamefont
  {Blackburn}}, \bibinfo {author} {\bibfnamefont {R.}~\bibnamefont {Caciuffo}},
  \bibinfo {author} {\bibfnamefont {N.}~\bibnamefont {Magnani}}, \bibinfo
  {author} {\bibfnamefont {P.}~\bibnamefont {Santini}}, \bibinfo {author}
  {\bibfnamefont {P.~J.}\ \bibnamefont {Brown}}, \bibinfo {author}
  {\bibfnamefont {M.}~\bibnamefont {Enderle}}, \ and\ \bibinfo {author}
  {\bibfnamefont {G.~H.}\ \bibnamefont {Lander}},\ }\href@noop {} {\bibfield
  {journal} {\bibinfo  {journal} {Phys. Rev. B}\ }\textbf {\bibinfo {volume}
  {72}},\ \bibinfo {pages} {184411} (\bibinfo {year} {2005})}\BibitemShut
  {NoStop}%
\bibitem [{\citenamefont {Wilkins}\ \emph {et~al.}(2006)\citenamefont
  {Wilkins}, \citenamefont {Caciuffo}, \citenamefont {Detlefs}, \citenamefont
  {Rebizant}, \citenamefont {Colineau}, \citenamefont {Wastin},\ and\
  \citenamefont {Lander}}]{Wilkins2006PRB60406}%
  \BibitemOpen
  \bibfield  {author} {\bibinfo {author} {\bibfnamefont {S.~B.}\ \bibnamefont
  {Wilkins}}, \bibinfo {author} {\bibfnamefont {R.}~\bibnamefont {Caciuffo}},
  \bibinfo {author} {\bibfnamefont {C.}~\bibnamefont {Detlefs}}, \bibinfo
  {author} {\bibfnamefont {J.}~\bibnamefont {Rebizant}}, \bibinfo {author}
  {\bibfnamefont {E.}~\bibnamefont {Colineau}}, \bibinfo {author}
  {\bibfnamefont {F.}~\bibnamefont {Wastin}}, \ and\ \bibinfo {author}
  {\bibfnamefont {G.~H.}\ \bibnamefont {Lander}},\ }\href@noop {} {\bibfield
  {journal} {\bibinfo  {journal} {Phys. Rev. B}\ }\textbf {\bibinfo {volume}
  {73}},\ \bibinfo {pages} {060406} (\bibinfo {year} {2006})}\BibitemShut
  {NoStop}%
\bibitem [{\citenamefont {Santini}\ \emph {et~al.}(2009)\citenamefont
  {Santini}, \citenamefont {Carretta}, \citenamefont {Amoretti}, \citenamefont
  {Caciuffo}, \citenamefont {Magnani},\ and\ \citenamefont
  {Lander}}]{Santini2009RMP807}%
  \BibitemOpen
  \bibfield  {author} {\bibinfo {author} {\bibfnamefont {P.}~\bibnamefont
  {Santini}}, \bibinfo {author} {\bibfnamefont {S.}~\bibnamefont {Carretta}},
  \bibinfo {author} {\bibfnamefont {G.}~\bibnamefont {Amoretti}}, \bibinfo
  {author} {\bibfnamefont {R.}~\bibnamefont {Caciuffo}}, \bibinfo {author}
  {\bibfnamefont {N.}~\bibnamefont {Magnani}}, \ and\ \bibinfo {author}
  {\bibfnamefont {G.~H.}\ \bibnamefont {Lander}},\ }\href@noop {} {\bibfield
  {journal} {\bibinfo  {journal} {Rev. Mod. Phys.}\ }\textbf {\bibinfo {volume}
  {81}},\ \bibinfo {pages} {807} (\bibinfo {year} {2009})}\BibitemShut
  {NoStop}%
\bibitem [{\citenamefont {Allen}(1968{\natexlab{a}})}]{Allen1968PR530}%
  \BibitemOpen
  \bibfield  {author} {\bibinfo {author} {\bibfnamefont {S.~J.}\ \bibnamefont
  {Allen}},\ }\href@noop {} {\bibfield  {journal} {\bibinfo  {journal} {Phys.
  Rev.}\ }\textbf {\bibinfo {volume} {166}},\ \bibinfo {pages} {530} (\bibinfo
  {year} {1968}{\natexlab{a}})}\BibitemShut {NoStop}%
\bibitem [{\citenamefont {Allen}(1968{\natexlab{b}})}]{Allen1968PR492}%
  \BibitemOpen
  \bibfield  {author} {\bibinfo {author} {\bibfnamefont {S.~J.}\ \bibnamefont
  {Allen}},\ }\href@noop {} {\bibfield  {journal} {\bibinfo  {journal} {Phys.
  Rev.}\ }\textbf {\bibinfo {volume} {167}},\ \bibinfo {pages} {492} (\bibinfo
  {year} {1968}{\natexlab{b}})}\BibitemShut {NoStop}%
\bibitem [{\citenamefont {Rahman}\ and\ \citenamefont
  {Runciman}(1966)}]{Rahman1966JPCS1833}%
  \BibitemOpen
  \bibfield  {author} {\bibinfo {author} {\bibfnamefont {H.~U.}\ \bibnamefont
  {Rahman}}\ and\ \bibinfo {author} {\bibfnamefont {W.~A.}\ \bibnamefont
  {Runciman}},\ }\href@noop {} {\bibfield  {journal} {\bibinfo  {journal} {J.
  Phys. Chem. Solids}\ }\textbf {\bibinfo {volume} {27}},\ \bibinfo {pages}
  {1833} (\bibinfo {year} {1966})}\BibitemShut {NoStop}%
\bibitem [{\citenamefont {Rahman}(1998)}]{Rahman1998PLA306}%
  \BibitemOpen
  \bibfield  {author} {\bibinfo {author} {\bibfnamefont {H.~U.}\ \bibnamefont
  {Rahman}},\ }\href@noop {} {\bibfield  {journal} {\bibinfo  {journal} {Phys.
  Lett. A}\ }\textbf {\bibinfo {volume} {240}},\ \bibinfo {pages} {306}
  (\bibinfo {year} {1998})}\BibitemShut {NoStop}%
\bibitem [{\citenamefont {Gajek}\ \emph {et~al.}(1988)\citenamefont {Gajek},
  \citenamefont {LAHALLE}, \citenamefont {Krupa},\ and\ \citenamefont
  {MULAK}}]{Gajek1988JLM351}%
  \BibitemOpen
  \bibfield  {author} {\bibinfo {author} {\bibfnamefont {Z.}~\bibnamefont
  {Gajek}}, \bibinfo {author} {\bibfnamefont {M.~P.}\ \bibnamefont {LAHALLE}},
  \bibinfo {author} {\bibfnamefont {J.~C.}\ \bibnamefont {Krupa}}, \ and\
  \bibinfo {author} {\bibfnamefont {J.}~\bibnamefont {MULAK}},\ }\href@noop {}
  {\bibfield  {journal} {\bibinfo  {journal} {J. Less-Common. Met.}\ }\textbf
  {\bibinfo {volume} {139}},\ \bibinfo {pages} {351} (\bibinfo {year}
  {1988})}\BibitemShut {NoStop}%
\bibitem [{\citenamefont {Magnani}\ \emph {et~al.}(2005)\citenamefont
  {Magnani}, \citenamefont {Santini}, \citenamefont {Amoretti},\ and\
  \citenamefont {Caciuffo}}]{Magnani2005PRB54405}%
  \BibitemOpen
  \bibfield  {author} {\bibinfo {author} {\bibfnamefont {N.}~\bibnamefont
  {Magnani}}, \bibinfo {author} {\bibfnamefont {P.}~\bibnamefont {Santini}},
  \bibinfo {author} {\bibfnamefont {G.}~\bibnamefont {Amoretti}}, \ and\
  \bibinfo {author} {\bibfnamefont {R.}~\bibnamefont {Caciuffo}},\ }\href@noop
  {} {\bibfield  {journal} {\bibinfo  {journal} {Phys. Rev. B}\ }\textbf
  {\bibinfo {volume} {71}},\ \bibinfo {pages} {054405} (\bibinfo {year}
  {2005})}\BibitemShut {NoStop}%
\bibitem [{\citenamefont {Kudin}\ \emph {et~al.}(2002)\citenamefont {Kudin},
  \citenamefont {Scuseria},\ and\ \citenamefont {Martin}}]{Kudin2002PRL266402}%
  \BibitemOpen
  \bibfield  {author} {\bibinfo {author} {\bibfnamefont {K.~N.}\ \bibnamefont
  {Kudin}}, \bibinfo {author} {\bibfnamefont {G.~E.}\ \bibnamefont {Scuseria}},
  \ and\ \bibinfo {author} {\bibfnamefont {R.~L.}\ \bibnamefont {Martin}},\
  }\href@noop {} {\bibfield  {journal} {\bibinfo  {journal} {Phys. Rev. Lett.}\
  }\textbf {\bibinfo {volume} {89}},\ \bibinfo {pages} {266402} (\bibinfo
  {year} {2002})}\BibitemShut {NoStop}%
\bibitem [{\citenamefont {Prodan}\ \emph {et~al.}(2007)\citenamefont {Prodan},
  \citenamefont {Scuseria},\ and\ \citenamefont {Martin}}]{Prodan2007PRB33101}%
  \BibitemOpen
  \bibfield  {author} {\bibinfo {author} {\bibfnamefont {I.~D.}\ \bibnamefont
  {Prodan}}, \bibinfo {author} {\bibfnamefont {G.~E.}\ \bibnamefont
  {Scuseria}}, \ and\ \bibinfo {author} {\bibfnamefont {R.~L.}\ \bibnamefont
  {Martin}},\ }\href@noop {} {\bibfield  {journal} {\bibinfo  {journal} {Phys.
  Rev. B}\ }\textbf {\bibinfo {volume} {76}},\ \bibinfo {pages} {033101}
  (\bibinfo {year} {2007})}\BibitemShut {NoStop}%
\bibitem [{\citenamefont {Becke}(1993)}]{Becke1993JCP1372}%
  \BibitemOpen
  \bibfield  {author} {\bibinfo {author} {\bibfnamefont {A.~D.}\ \bibnamefont
  {Becke}},\ }\href@noop {} {\bibfield  {journal} {\bibinfo  {journal} {J.
  Chem. Phys.}\ }\textbf {\bibinfo {volume} {98}},\ \bibinfo {pages} {1372}
  (\bibinfo {year} {1993})}\BibitemShut {NoStop}%
\bibitem [{\citenamefont {Petit}\ \emph {et~al.}(2010)\citenamefont {Petit},
  \citenamefont {Svane}, \citenamefont {Szotek}, \citenamefont {Temmerman},\
  and\ \citenamefont {Stocks}}]{Petit2010PRB45108}%
  \BibitemOpen
  \bibfield  {author} {\bibinfo {author} {\bibfnamefont {L.}~\bibnamefont
  {Petit}}, \bibinfo {author} {\bibfnamefont {A.}~\bibnamefont {Svane}},
  \bibinfo {author} {\bibfnamefont {Z.}~\bibnamefont {Szotek}}, \bibinfo
  {author} {\bibfnamefont {W.~M.}\ \bibnamefont {Temmerman}}, \ and\ \bibinfo
  {author} {\bibfnamefont {G.}~\bibnamefont {Stocks}},\ }\href@noop {}
  {\bibfield  {journal} {\bibinfo  {journal} {Phys. Rev. B}\ }\textbf {\bibinfo
  {volume} {81}},\ \bibinfo {pages} {045108} (\bibinfo {year}
  {2010})}\BibitemShut {NoStop}%
\bibitem [{\citenamefont {Perdew}\ and\ \citenamefont
  {Zunger}(1981)}]{Perdew1981PRB5048}%
  \BibitemOpen
  \bibfield  {author} {\bibinfo {author} {\bibfnamefont {J.~P.}\ \bibnamefont
  {Perdew}}\ and\ \bibinfo {author} {\bibfnamefont {A.}~\bibnamefont
  {Zunger}},\ }\href@noop {} {\bibfield  {journal} {\bibinfo  {journal} {Phys.
  Rev. B}\ }\textbf {\bibinfo {volume} {23}},\ \bibinfo {pages} {5048}
  (\bibinfo {year} {1981})}\BibitemShut {NoStop}%
\bibitem [{\citenamefont {Dudarev}\ \emph {et~al.}(1997)\citenamefont
  {Dudarev}, \citenamefont {Manh},\ and\ \citenamefont
  {Sutton}}]{Dudarev1997PMB613}%
  \BibitemOpen
  \bibfield  {author} {\bibinfo {author} {\bibfnamefont {S.~L.}\ \bibnamefont
  {Dudarev}}, \bibinfo {author} {\bibfnamefont {D.~N.}\ \bibnamefont {Manh}}, \
  and\ \bibinfo {author} {\bibfnamefont {A.~P.}\ \bibnamefont {Sutton}},\
  }\href@noop {} {\bibfield  {journal} {\bibinfo  {journal} {Philos. Mag. B}\
  }\textbf {\bibinfo {volume} {75}},\ \bibinfo {pages} {613} (\bibinfo {year}
  {1997})}\BibitemShut {NoStop}%
\bibitem [{\citenamefont {Anisimov}\ \emph {et~al.}(1991)\citenamefont
  {Anisimov}, \citenamefont {Zaanen},\ and\ \citenamefont
  {Andersen}}]{Anisimov1991PRB943}%
  \BibitemOpen
  \bibfield  {author} {\bibinfo {author} {\bibfnamefont {V.~I.}\ \bibnamefont
  {Anisimov}}, \bibinfo {author} {\bibfnamefont {J.}~\bibnamefont {Zaanen}}, \
  and\ \bibinfo {author} {\bibfnamefont {O.~K.}\ \bibnamefont {Andersen}},\
  }\href@noop {} {\bibfield  {journal} {\bibinfo  {journal} {Phys. Rev. B}\
  }\textbf {\bibinfo {volume} {44}},\ \bibinfo {pages} {943} (\bibinfo {year}
  {1991})}\BibitemShut {NoStop}%
\bibitem [{\citenamefont {Colarieti-Tosti}\ \emph {et~al.}(2002)\citenamefont
  {Colarieti-Tosti}, \citenamefont {Eriksson}, \citenamefont {Nordstrom},
  \citenamefont {Wills},\ and\ \citenamefont
  {Brooks}}]{Colarieti-Tosti2002PRB195102}%
  \BibitemOpen
  \bibfield  {author} {\bibinfo {author} {\bibfnamefont {M.}~\bibnamefont
  {Colarieti-Tosti}}, \bibinfo {author} {\bibfnamefont {O.}~\bibnamefont
  {Eriksson}}, \bibinfo {author} {\bibfnamefont {L.}~\bibnamefont {Nordstrom}},
  \bibinfo {author} {\bibfnamefont {J.}~\bibnamefont {Wills}}, \ and\ \bibinfo
  {author} {\bibfnamefont {M.~S.~S.}\ \bibnamefont {Brooks}},\ }\href@noop {}
  {\bibfield  {journal} {\bibinfo  {journal} {Phys. Rev. B}\ }\textbf {\bibinfo
  {volume} {65}},\ \bibinfo {pages} {195102} (\bibinfo {year}
  {2002})}\BibitemShut {NoStop}%
\bibitem [{\citenamefont {Novak}\ and\ \citenamefont
  {Divis}(2007)}]{Novak2007PSSB3168}%
  \BibitemOpen
  \bibfield  {author} {\bibinfo {author} {\bibfnamefont {P.}~\bibnamefont
  {Novak}}\ and\ \bibinfo {author} {\bibfnamefont {M.}~\bibnamefont {Divis}},\
  }\href@noop {} {\bibfield  {journal} {\bibinfo  {journal} {Phys. Status
  Solidi B}\ }\textbf {\bibinfo {volume} {244}},\ \bibinfo {pages} {3168}
  (\bibinfo {year} {2007})}\BibitemShut {NoStop}%
\bibitem [{\citenamefont {Laskowski}\ \emph {et~al.}(2004)\citenamefont
  {Laskowski}, \citenamefont {Madsen}, \citenamefont {Blaha},\ and\
  \citenamefont {Schwarz}}]{Laskowski2004PRB140408}%
  \BibitemOpen
  \bibfield  {author} {\bibinfo {author} {\bibfnamefont {R.}~\bibnamefont
  {Laskowski}}, \bibinfo {author} {\bibfnamefont {G.~K.~H.}\ \bibnamefont
  {Madsen}}, \bibinfo {author} {\bibfnamefont {P.}~\bibnamefont {Blaha}}, \
  and\ \bibinfo {author} {\bibfnamefont {K.}~\bibnamefont {Schwarz}},\
  }\href@noop {} {\bibfield  {journal} {\bibinfo  {journal} {Phys. Rev. B}\
  }\textbf {\bibinfo {volume} {69}},\ \bibinfo {pages} {140408} (\bibinfo
  {year} {2004})}\BibitemShut {NoStop}%
\bibitem [{\citenamefont {Zhou}\ and\ \citenamefont
  {Ozolins}(2009)}]{Zhou2009PRB125127}%
  \BibitemOpen
  \bibfield  {author} {\bibinfo {author} {\bibfnamefont {F.}~\bibnamefont
  {Zhou}}\ and\ \bibinfo {author} {\bibfnamefont {V.}~\bibnamefont {Ozolins}},\
  }\href@noop {} {\bibfield  {journal} {\bibinfo  {journal} {Phys. Rev. B}\
  }\textbf {\bibinfo {volume} {80}},\ \bibinfo {pages} {125127} (\bibinfo
  {year} {2009})}\BibitemShut {NoStop}%
\bibitem [{\citenamefont {Shick}\ \emph {et~al.}(2001)\citenamefont {Shick},
  \citenamefont {Pickett},\ and\ \citenamefont
  {Liechtenstein}}]{Shick2001JES753}%
  \BibitemOpen
  \bibfield  {author} {\bibinfo {author} {\bibfnamefont {A.~B.}\ \bibnamefont
  {Shick}}, \bibinfo {author} {\bibfnamefont {W.~E.}\ \bibnamefont {Pickett}},
  \ and\ \bibinfo {author} {\bibfnamefont {A.~I.}\ \bibnamefont
  {Liechtenstein}},\ }\href@noop {} {\bibfield  {journal} {\bibinfo  {journal}
  {J. Electron Spectrosc.}\ }\textbf {\bibinfo {volume} {114-116}},\ \bibinfo
  {pages} {753} (\bibinfo {year} {2001})}\BibitemShut {NoStop}%
\bibitem [{\citenamefont {Larson}\ \emph {et~al.}(2007)\citenamefont {Larson},
  \citenamefont {Lambrecht}, \citenamefont {Chantis},\ and\ \citenamefont {van
  Schilfgaarde}}]{Larson2007PRB45114}%
  \BibitemOpen
  \bibfield  {author} {\bibinfo {author} {\bibfnamefont {P.}~\bibnamefont
  {Larson}}, \bibinfo {author} {\bibfnamefont {W.~R.~L.}\ \bibnamefont
  {Lambrecht}}, \bibinfo {author} {\bibfnamefont {A.}~\bibnamefont {Chantis}},
  \ and\ \bibinfo {author} {\bibfnamefont {M.}~\bibnamefont {van
  Schilfgaarde}},\ }\href@noop {} {\bibfield  {journal} {\bibinfo  {journal}
  {Phys. Rev. B}\ }\textbf {\bibinfo {volume} {75}},\ \bibinfo {pages} {045114}
  (\bibinfo {year} {2007})}\BibitemShut {NoStop}%
\bibitem [{\citenamefont {Jomard}\ \emph {et~al.}(2008)\citenamefont {Jomard},
  \citenamefont {Amadon}, \citenamefont {Bottin},\ and\ \citenamefont
  {Torrent}}]{Jomard2008PRB75125}%
  \BibitemOpen
  \bibfield  {author} {\bibinfo {author} {\bibfnamefont {G.}~\bibnamefont
  {Jomard}}, \bibinfo {author} {\bibfnamefont {B.}~\bibnamefont {Amadon}},
  \bibinfo {author} {\bibfnamefont {F.}~\bibnamefont {Bottin}}, \ and\ \bibinfo
  {author} {\bibfnamefont {M.}~\bibnamefont {Torrent}},\ }\href@noop {}
  {\bibfield  {journal} {\bibinfo  {journal} {Phys. Rev. B}\ }\textbf {\bibinfo
  {volume} {78}},\ \bibinfo {pages} {075125} (\bibinfo {year}
  {2008})}\BibitemShut {NoStop}%
\bibitem [{\citenamefont {Amadon}\ \emph {et~al.}(2008)\citenamefont {Amadon},
  \citenamefont {Jollet},\ and\ \citenamefont {Torrent}}]{Amadon2008PRB155104}%
  \BibitemOpen
  \bibfield  {author} {\bibinfo {author} {\bibfnamefont {B.}~\bibnamefont
  {Amadon}}, \bibinfo {author} {\bibfnamefont {F.}~\bibnamefont {Jollet}}, \
  and\ \bibinfo {author} {\bibfnamefont {M.}~\bibnamefont {Torrent}},\
  }\href@noop {} {\bibfield  {journal} {\bibinfo  {journal} {Phys. Rev. B}\
  }\textbf {\bibinfo {volume} {77}},\ \bibinfo {pages} {155104} (\bibinfo
  {year} {2008})}\BibitemShut {NoStop}%
\bibitem [{\citenamefont {Ylvisaker}\ \emph {et~al.}(2009)\citenamefont
  {Ylvisaker}, \citenamefont {Pickett},\ and\ \citenamefont
  {Koepernik}}]{Ylvisaker2009PRB35103}%
  \BibitemOpen
  \bibfield  {author} {\bibinfo {author} {\bibfnamefont {E.~R.}\ \bibnamefont
  {Ylvisaker}}, \bibinfo {author} {\bibfnamefont {W.~E.}\ \bibnamefont
  {Pickett}}, \ and\ \bibinfo {author} {\bibfnamefont {K.}~\bibnamefont
  {Koepernik}},\ }\href@noop {} {\bibfield  {journal} {\bibinfo  {journal}
  {Phys. Rev. B}\ }\textbf {\bibinfo {volume} {79}},\ \bibinfo {pages} {035103}
  (\bibinfo {year} {2009})}\BibitemShut {NoStop}%
\bibitem [{\citenamefont {Dorado}\ \emph {et~al.}(2009)\citenamefont {Dorado},
  \citenamefont {Amadon}, \citenamefont {Freyss},\ and\ \citenamefont
  {Bertolus}}]{Dorado2009PRB235125}%
  \BibitemOpen
  \bibfield  {author} {\bibinfo {author} {\bibfnamefont {B.}~\bibnamefont
  {Dorado}}, \bibinfo {author} {\bibfnamefont {B.}~\bibnamefont {Amadon}},
  \bibinfo {author} {\bibfnamefont {M.}~\bibnamefont {Freyss}}, \ and\ \bibinfo
  {author} {\bibfnamefont {M.}~\bibnamefont {Bertolus}},\ }\href@noop {}
  {\bibfield  {journal} {\bibinfo  {journal} {Phys. Rev. B}\ }\textbf {\bibinfo
  {volume} {79}},\ \bibinfo {pages} {235125} (\bibinfo {year}
  {2009})}\BibitemShut {NoStop}%
\bibitem [{\citenamefont {Meredig}\ \emph {et~al.}(2010)\citenamefont
  {Meredig}, \citenamefont {Thompson}, \citenamefont {Hansen}, \citenamefont
  {Wolverton},\ and\ \citenamefont {van~de Walle}}]{Meredig2010PRB195128}%
  \BibitemOpen
  \bibfield  {author} {\bibinfo {author} {\bibfnamefont {B.}~\bibnamefont
  {Meredig}}, \bibinfo {author} {\bibfnamefont {A.}~\bibnamefont {Thompson}},
  \bibinfo {author} {\bibfnamefont {H.~A.}\ \bibnamefont {Hansen}}, \bibinfo
  {author} {\bibfnamefont {C.}~\bibnamefont {Wolverton}}, \ and\ \bibinfo
  {author} {\bibfnamefont {A.}~\bibnamefont {van~de Walle}},\ }\href@noop {}
  {\bibfield  {journal} {\bibinfo  {journal} {Phys. Rev. B}\ }\textbf {\bibinfo
  {volume} {82}},\ \bibinfo {pages} {195128} (\bibinfo {year}
  {2010})}\BibitemShut {NoStop}%
\bibitem [{\citenamefont {Kresse}\ and\ \citenamefont
  {Joubert}(1999)}]{Kresse1999PRB1758}%
  \BibitemOpen
  \bibfield  {author} {\bibinfo {author} {\bibfnamefont {G.}~\bibnamefont
  {Kresse}}\ and\ \bibinfo {author} {\bibfnamefont {D.}~\bibnamefont
  {Joubert}},\ }\href@noop {} {\bibfield  {journal} {\bibinfo  {journal} {Phys.
  Rev. B}\ }\textbf {\bibinfo {volume} {59}},\ \bibinfo {pages} {1758}
  (\bibinfo {year} {1999})}\BibitemShut {NoStop}%
\bibitem [{\citenamefont {Blochl}(1994)}]{Blochl1994PRB17953}%
  \BibitemOpen
  \bibfield  {author} {\bibinfo {author} {\bibfnamefont {P.~E.}\ \bibnamefont
  {Blochl}},\ }\href@noop {} {\bibfield  {journal} {\bibinfo  {journal} {Phys.
  Rev. B}\ }\textbf {\bibinfo {volume} {50}},\ \bibinfo {pages} {17953}
  (\bibinfo {year} {1994})}\BibitemShut {NoStop}%
\bibitem [{\citenamefont {Liechtenstein}\ \emph {et~al.}(1995)\citenamefont
  {Liechtenstein}, \citenamefont {Anisimov},\ and\ \citenamefont
  {Zaanen}}]{Liechtenstein1995PRB5467}%
  \BibitemOpen
  \bibfield  {author} {\bibinfo {author} {\bibfnamefont {A.~I.}\ \bibnamefont
  {Liechtenstein}}, \bibinfo {author} {\bibfnamefont {V.~I.}\ \bibnamefont
  {Anisimov}}, \ and\ \bibinfo {author} {\bibfnamefont {J.}~\bibnamefont
  {Zaanen}},\ }\href@noop {} {\bibfield  {journal} {\bibinfo  {journal} {Phys.
  Rev. B}\ }\textbf {\bibinfo {volume} {52}},\ \bibinfo {pages} {R5467}
  (\bibinfo {year} {1995})}\BibitemShut {NoStop}%
\bibitem [{\citenamefont {Judd}(1963)}]{Judd1963}%
  \BibitemOpen
  \bibfield  {author} {\bibinfo {author} {\bibfnamefont {B.~R.}\ \bibnamefont
  {Judd}},\ }\href@noop {} {\emph {\bibinfo {title} {Operator Techniques in
  Atomic Spectroscopy}}}\ (\bibinfo  {publisher} {McGraw-Hill},\ \bibinfo
  {address} {New York},\ \bibinfo {year} {1963})\BibitemShut {NoStop}%
\bibitem [{\citenamefont {Newman}\ and\ \citenamefont {Ng}(2000)}]{Newman2000}%
  \BibitemOpen
  \bibfield  {author} {\bibinfo {author} {\bibfnamefont {D.}~\bibnamefont
  {Newman}}\ and\ \bibinfo {author} {\bibfnamefont {B.}~\bibnamefont {Ng}},\
  }\href@noop {} {\emph {\bibinfo {title} {Crystal Field Handbook}}}\ (\bibinfo
   {publisher} {Cambridge University Press},\ \bibinfo {address} {Cambridge},\
  \bibinfo {year} {2000})\BibitemShut {NoStop}%
\bibitem [{\citenamefont {Carnall}(1992)}]{Carnall1992JCP8713}%
  \BibitemOpen
  \bibfield  {author} {\bibinfo {author} {\bibfnamefont {W.~T.}\ \bibnamefont
  {Carnall}},\ }\href@noop {} {\bibfield  {journal} {\bibinfo  {journal} {J.
  Chem. Phys.}\ }\textbf {\bibinfo {volume} {96}},\ \bibinfo {pages} {8713}
  (\bibinfo {year} {1992})}\BibitemShut {NoStop}%
\bibitem [{\citenamefont {Berry}\ \emph {et~al.}(1988)\citenamefont {Berry},
  \citenamefont {Schwieters},\ and\ \citenamefont
  {Richardson}}]{Berry1988CP105}%
  \BibitemOpen
  \bibfield  {author} {\bibinfo {author} {\bibfnamefont {M.~T.}\ \bibnamefont
  {Berry}}, \bibinfo {author} {\bibfnamefont {C.}~\bibnamefont {Schwieters}}, \
  and\ \bibinfo {author} {\bibfnamefont {F.~S.}\ \bibnamefont {Richardson}},\
  }\href@noop {} {\bibfield  {journal} {\bibinfo  {journal} {Chem. Phys.}\
  }\textbf {\bibinfo {volume} {122}},\ \bibinfo {pages} {105} (\bibinfo {year}
  {1988})}\BibitemShut {NoStop}%
\bibitem [{\citenamefont {Van~Deurzen}\ \emph {et~al.}(1984)\citenamefont
  {Van~Deurzen}, \citenamefont {Rajnak},\ and\ \citenamefont
  {Conway}}]{VanDeurzen1984JOSAB45}%
  \BibitemOpen
  \bibfield  {author} {\bibinfo {author} {\bibfnamefont {C.~H.~H.}\
  \bibnamefont {Van~Deurzen}}, \bibinfo {author} {\bibfnamefont
  {K.}~\bibnamefont {Rajnak}}, \ and\ \bibinfo {author} {\bibfnamefont {J.~G.}\
  \bibnamefont {Conway}},\ }\href@noop {} {\bibfield  {journal} {\bibinfo
  {journal} {J. Opt. Soc. Am. B}\ }\textbf {\bibinfo {volume} {1}},\ \bibinfo
  {pages} {45} (\bibinfo {year} {1984})}\BibitemShut {NoStop}%
\bibitem [{\citenamefont {Liu}(2005)}]{Liu2005JSSC489}%
  \BibitemOpen
  \bibfield  {author} {\bibinfo {author} {\bibfnamefont {G.~K.}\ \bibnamefont
  {Liu}},\ }\href@noop {} {\bibfield  {journal} {\bibinfo  {journal} {J. Solid
  State Chem.}\ }\textbf {\bibinfo {volume} {178}},\ \bibinfo {pages} {489}
  (\bibinfo {year} {2005})}\BibitemShut {NoStop}%
\bibitem [{\citenamefont {Ippolito}\ \emph {et~al.}(2005)\citenamefont
  {Ippolito}, \citenamefont {Martinelli},\ and\ \citenamefont
  {Bevilacqua}}]{Ippolito2005PRB64419}%
  \BibitemOpen
  \bibfield  {author} {\bibinfo {author} {\bibfnamefont {D.}~\bibnamefont
  {Ippolito}}, \bibinfo {author} {\bibfnamefont {L.}~\bibnamefont
  {Martinelli}}, \ and\ \bibinfo {author} {\bibfnamefont {G.}~\bibnamefont
  {Bevilacqua}},\ }\href@noop {} {\bibfield  {journal} {\bibinfo  {journal}
  {Phys. Rev. B}\ }\textbf {\bibinfo {volume} {71}},\ \bibinfo {pages} {064419}
  (\bibinfo {year} {2005})}\BibitemShut {NoStop}%
\bibitem [{\citenamefont {McKenzie}\ \emph {et~al.}(1982)\citenamefont
  {McKenzie}, \citenamefont {Domb},\ and\ \citenamefont
  {Hunter}}]{McKenzie1982JPA3899}%
  \BibitemOpen
  \bibfield  {author} {\bibinfo {author} {\bibfnamefont {S.}~\bibnamefont
  {McKenzie}}, \bibinfo {author} {\bibfnamefont {C.}~\bibnamefont {Domb}}, \
  and\ \bibinfo {author} {\bibfnamefont {D.~L.}\ \bibnamefont {Hunter}},\
  }\href@noop {} {\bibfield  {journal} {\bibinfo  {journal} {J. Phys. A}\
  }\textbf {\bibinfo {volume} {15}},\ \bibinfo {pages} {3899} (\bibinfo {year}
  {1982})}\BibitemShut {NoStop}%
\end{thebibliography}

\end{document}